\documentclass[twocolumn]{bmcart}

\usepackage{amsthm,amsmath}
\usepackage[utf8]{inputenc} 

\usepackage{graphicx}

\startlocaldefs
\endlocaldefs

\begin{document}

\begin{frontmatter}

\begin{fmbox}
\dochead{Research}


\title{Deep Ad-hoc Beamforming Based on Speaker Extraction for Target-Dependent Speech Separation}


\author[
  addressref={aff1},                   
  email={2015300797@mail.nwpu.edu.cn}   
]{\inits{Z.Y.Y}\fnm{Ziye} \snm{Yang}}
\author[
  addressref={aff1},
  email={gshanzheng@mail.nwpu.edu.cn}
]{\inits{S.Z.G}\fnm{Shanzheng} \snm{Guan}}
\author[
  addressref={aff1},
  corref={aff1},
  email={xiaolei.zhang@nwpu.edu.cn}
]{\inits{X.L.Z}\fnm{Xiao-Lei} \snm{Zhang}}


\address[id=aff1]{
  \orgdiv{1Center for Intelligent Acoustics and Immersive Communications, School of Marine Science and Technology},             
  \orgname{ Northwestern Polytechnical University},          
  \city{Xi'an},                              
  \cny{China}                                    
}



\end{fmbox}


\begin{abstractbox}

\begin{abstract} 
Recently, the research on ad-hoc microphone arrays with deep learning has drawn much attention, especially in speech enhancement and separation. Because an ad-hoc microphone array may cover such a large area that multiple speakers may locate far apart and talk independently, target-dependent speech separation, which aims to extract a target speaker from a mixed speech, is important for extracting and tracing a specific speaker in the ad-hoc array. However, this technique has not been explored yet. In this paper, we propose deep ad-hoc beamforming based on speaker extraction, which is to our knowledge the first work for target-dependent speech separation based on ad-hoc microphone arrays and deep learning. The algorithm contains three components. First, we propose a supervised channel selection framework based on speaker extraction, where the estimated utterance-level SNRs of the target speech are used as the basis for the channel selection. Second, we apply the selected channels to a deep learning based MVDR algorithm, where a single-channel speaker extraction algorithm is applied to each selected channel for estimating the mask of the target speech. We conducted an extensive experiment on a WSJ0-adhoc corpus. Experimental results demonstrate the effectiveness of the proposed method.

\end{abstract}




\end{abstractbox}
%

\end{frontmatter}



\section{Introduction}
  Speech separation, also known as cocktail party problem, aims to separate target speech from interference background \cite{wang2018supervised}. It is often used as the front end of speech recognition for improving the accuracy of human-machine interaction. Conventional speech separation technologies include computational auditory scene analysis \cite{rouat2008computational}, non-negative matrix factorization \cite{schmidt2006single,virtanen2007monaural}, HMM-GMM \cite{virtanen2006speech,stark2010source}, and minimum mean square error \cite{ephraim1985speech}. Recently, deep learning based speech separation becomes a new trend \cite{wang2012boosting,wang2013towards,wang2014training,zhang2017deep,delfarah2019deep,delfarah2020two}, which is the focus of this paper. According to whether speakers’ information is known as a prior, deep-learning-based speech separation techniques can be divided into three categories, which are speaker-dependent \cite{bregman1994auditory}, target-dependent, and speaker-independent speech separation.
   Speaker-dependent speech separation needs to known the prior information of all speakers, which limits its practical applications. Nowadays, the research on speech separation is mostly speaker-independent and target-dependent.

    Speaker-independent speech separation based on deep learning faces the speaker permutation ambiguity problem. In order to solve this problem, two techniques have been proposed. The first one is deep clustering 
    \cite{hershey2016deep,wang2018alternative,yang2019multi}.
    It projects each time-frequency unit to a higher-dimensional embedding vector by a deep network, and conducts clustering on the embedding vectors for speech separation.
    The second technique is permutation invariant training 
    \cite{yu2017permutation,kolbaek2017multitalker,xu2018single}. For each training mixture, it picks the permutation of the speakers that has the minimum training error among all possible permutations to train the network.


Target-dependent speech separation based on deep learning aims to extract target speech from a mixture given some prior knowledge on the target speaker. The earliest speech separation method takes the target speaker as the training target \cite{zhang2016deep}. It has to train a model for each target speaker, which limits its practical use. To prevent training a model for each target speaker, \textit{speaker extraction} further takes speaker codes extracted from a speaker recognition system as part of the network input \cite{vzmolikova2017learning,xu2019optimization,xiao2019single,delcroix2019compact,ochiai2019unified,wang2018voicefilter,williamson2017time}.
Some representative speaker extraction methods are as follows. \cite{delcroix2018single} applies a context adaptive deep neural network to extract the target speaker through a speaker adaptation layer. It takes the estimated mask and ideal binary mask
as the training objective. \cite{xu2019optimization} proposed a temporal spectrum approximation loss to estimate a phase sensitive mask for the target speaker. \cite{9004016} generalized the end-to-end speaker-independent speech separation \cite{luo2019conv} to the end-to-end speaker extraction.
\begin{figure*}[t]
 \centering
  \includegraphics[width=0.9\textwidth]{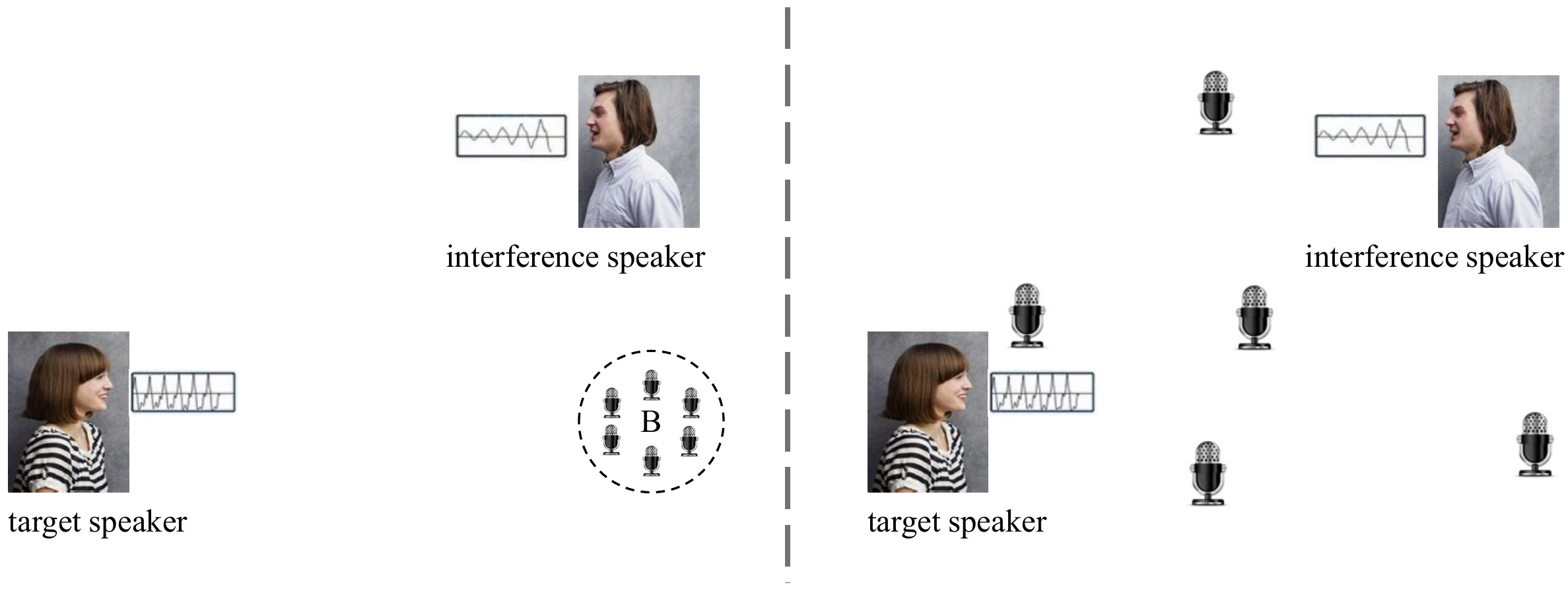}
  \caption{ The scenarios of speaker extraction under traditional microphone array and DABse.}
  \label{fig:1}
  \end{figure*}

The aforementioned methods are all single-channel methods. Although they work well in clean scenarios, their performance degrades significantly in reverberant scenarios. To improve the performance of speech separation in reverberant scenarios, many multichannel methods were proposed, which has the following two major forms. The first form combines spatial features that are extracted from microphone arrays, such as interaural time difference and interaural level difference, with spectral features as the input of single-channel speech separation networks \cite{wang2018multi,jiang2014binaural,araki2015exploring,pertila2015distant}. The second form uses a deep network to predict a mask for each speaker at each channel, and then conducts beamforming for each speaker \cite{nakatani2017integrating}. For brevity, we call this method \textit{deep beamforming}. Some methods combined the above two forms for boosting their advantages together in reverberant scenarios, e.g. \cite{yang2019boosting}.

The aforementioned multichannel methods are only studied with traditional fixed arrays, such as linear arrays or spherical arrays. However, for  far-field speech separation problems with high reverberation, they suffer significant performance degradation. How to maintain the estimated speech at the same high quality throughout an interested physical space is of broad interests.  Ad-hoc microphone array, which is a group of randomly distributed microphones collaborating with each other, is a solution to the problem. Figure \ref{fig:1} gives a comparison example where a target speaker extraction problem with a fixed array is on the left and that with an ad-hoc microphone array on the right.
From the figure, we see that, compared with the fixed array that is far from the target speaker, the ad-hoc microphone array has several apparent advantages. First, an ad-hoc microphone array may put a number of microphones around the target speaker, which significantly reduced the probability of far-field speech processing. By channel selection, it might be able to form a local microphone array around the target speaker. At last, it may be able to incorporate application devices of various physical sizes.

 \begin{figure*}[t]
    \includegraphics[width=0.9\textwidth]{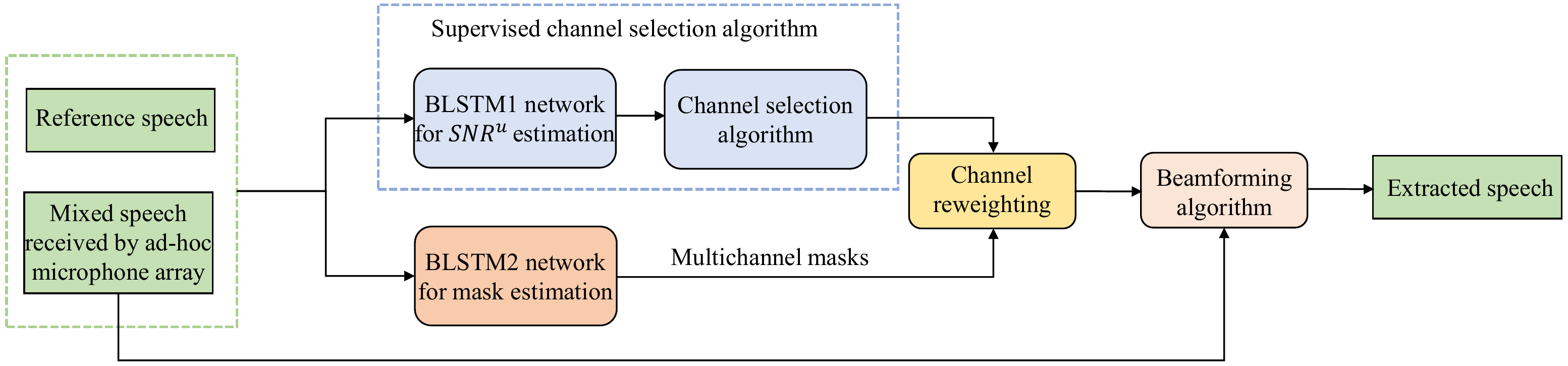}
  \caption{Diagram of the proposed DABse system.}
    \label{fig:2}
\end{figure*}

In literature, ad-hoc microphone arrays have consistently been an important research topic \cite{jayaprakasam2017distributed,tavakoli2017distributed,zhang2017microphone,koutrouvelis2018low}. However, they face many practical problems due to the lack of important priors. Recently, \cite{zhang2018deep} addresses the difficulties of ad-hoc microphone arrays, such as lack of priors and insufficient estimation of variables, by deep learning for the first time. The proposed method, named deep ad-hoc beamforming (DAB), was originally designed for speech enhancement only, which predicts segment-level signal-to-noise-ratio (SNR) by deep neural networks for supervised channel selection. Later on,
some speech separation methods based on ad-hoc microphone arrays were proposed. \cite{luo2020end} proposed a transform-average-concatenate strategy for a filter-and-sum network \cite{luo2019fasnet} to realize the channel reweighting/selection ability for ad-hoc microphone arrays. Because ad-hoc microphone arrays lack the prior of the number and spatial distribution of microphones, \cite{wang2020neural} proposed a network architecture by interleaving inter-channel processing layers and temporal processing layers to leverage information across time and space alternately. 


However, existing deep learning based speech separation with ad-hoc microphone arrays are all speaker-independent. To our knowledge, target-dependent speech separation with ad-hoc microphone arrays are far from explored yet. In many applications, extracting and tracking target speech is of more interests than separating a mixture into its components. This is particularly the case for ad-hoc microphone arrays, where several speakers may locate far apart and talk independently.

In this paper, we propose a target-dependent speech separation algorithm with ad-hoc microphone arrays, named DAB based on speaker extraction (DABse).
Our algorithm consists of three components: first, we propose a supervised channel selection based on speaker extraction, which applies bi-directional long short-term memory (BLSTM) networks to estimate the utterance-level SNR of the target speaker. Then, we employ the heuristic channel selection algorithms in \cite{zhang2018deep} to pick the channels with high SNRs. We further apply a single-channel speaker extraction algorithm to the selected channels for the mask estimation problem of the target speech. At last, we use the estimated masks to derive a beamformer for the target speaker, such as minimum variance distortionless response (MVDR) \cite{heymann2016neural}.
Experimental results on a WSJ0-adhoc corpus show that the proposed DABse performs well in reverberant environments.

The rest of the paper is organized as follows. We introduce the signal model of the speaker extraction problem with ad-hoc microphone arrays in Section \ref{sec:2}. Then, we present the deep ad-hoc beamforming system based on speaker extraction in Section \ref{sec:3}. In Section \ref{sec:4}, we present the experimental results. Finally, we conclude this study in Section \ref{sec:6}.

\section{Signal model}\label{sec:2}
  \begin{figure*}[t]
      \includegraphics[width=0.9\textwidth]{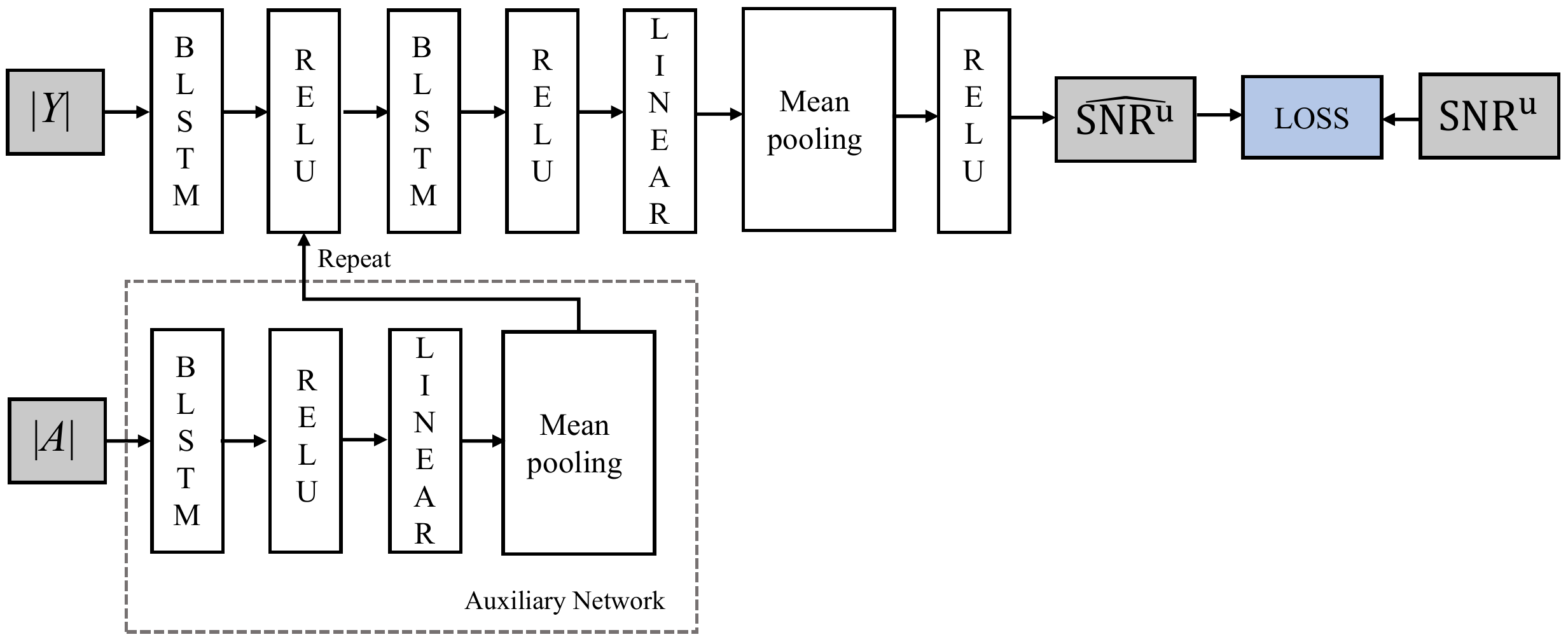}
  \caption{Diagram of the channel-weight estimation network.}
    \label{fig:3}
\end{figure*}
In this section, we build the signal model for target-dependent speech separation based on ad-hoc microphone arrays. Suppose that a room contains a target speaker, an interference speaker, and an ad-hoc array of $W$ microphones. Then, the mixed speech signal received by any single microphone of the ad-hoc array can be represented as:
\begin{equation}\label{eq:1}
y(t)= x_{a}(t)+x_{i}(t)+h(t)
\end{equation}
where $x_{a}(t)$ and $x_{i}(t)$ are the direct speech of the target speaker and interference speaker at time $t$, and $h(t)$ is the early and late reverberation of the speech source signal.

We perform the short-time Fourier transform (STFT) to the signal \eqref{eq:1}, which results in:
\begin{equation}
Y(t,f)= X_{a}(t,f)+X_{i}(t,f)+H(t,f)
\end{equation}
where $X_{a}(t,f)$ and $X_{i}(t,f)$ are the time-frequency units of the direct speech of the target speaker and interference speaker at time $t$ and frequency $f$ respectively, $H(t,f)$ is the time-frequency unit of the early and late reverberation. We can further define the direct speech as follows:
\begin{equation}
X_{a}(t,f) = c_{a}(f)S_{a}(t,f)
\end{equation}
\begin{equation}
X_{i}(t,f) = c_{i}(f)S_{i}(t,f)
\end{equation}
where $S_{a}(t,f)$ and $S_{i}(t,f)$ are the spectra of the target and interference speech at the source locations, and $c_{a}(f)$ and $c_{i}(f)$, which are complex numbers, are the time-invariant acoustic transfer functions from the speech sources to the microphone of the array.

\section{Deep ad-hoc beamforming based on speaker extraction}\label{sec:3}

Figure \ref{fig:2} describes DABse. It first picks eligible channels from ad-hoc microphone arrays by a supervised channel selection algorithm based on speaker extraction for a target speaker. Then, it conducts deep-learning-based MVDR on the selected channels, where a separate single-channel speaker extraction network is used to estimate the mask of the target speaker.

\subsection{Supervised channel selection based on speaker extraction}\label{sec:3.1}
The main idea of the supervised channel selection based on speaker extraction is to select the channels with high SNR of the target speaker. The module contains two parts: a channel-weight estimation network, and a channel selection algorithm.

\subsubsection{Channel-weight estimation network}

The channel-weight estimation network aims to estimate the quality of the target speech for each channel.
To make the channel-weight estimation network independent to the topology of ad-hoc microphone arrays, it needs to be trained in a single-channel fashion which is then applied to each channel separately in the test stage.
Here we use a speaker extraction network to estimate the quality of the target speech, where an auxiliary network is to extract the identity feature of the target speaker. Figure \ref{fig:3} shows the architecture of the channel-weight estimation network.

First of all, we need to define a training target. Many objective evaluation metrics are suitable to be used as the training target for evaluating the speech quality. As the first work of the target-dependent speech separation based on DAB, we take the simplest training target---a variant of SNR:
\begin{equation}
{\rm{SNR}}^u = \frac{\sum_{t} |x_a(t)|}{\sum_t |x_a(t)|+\sum_t|x_i(t)|}.
\end{equation}
 We name the variant of the SNR as the utterance-level SNR (${\rm{SNR}}^u$).

The network structure contains an auxiliary network and a main network. Suppose each target speaker has an auxiliary speech that is collected independently from the mixed speech $y(t)$. The auxiliary network takes the magnitude spectrum of the auxiliary speech $|{A}|$ as its input for extracting the identity embedding feature of the target speaker. It uses a BLSTM network to extract frame-level features from $|A|$, and then uses a pooling layer to transform the frame-level features into an utterance-level embedding vector $\mathbf{v}$.

The main network contains a frame-level network, a pooling layer, and an utterance-level network from bottom-up. The frame-level network first transforms $|Y|$ to frame-level features, denoted as $[\mathbf{z}_1,\ldots,\mathbf{z}_b,\ldots, \mathbf{z}_B]$, then concatenates each frame-level feature $\mathbf{z}_b$ with $\mathbf{v}$, and finally extracts a target-dependent frame-level feature from $[\mathbf{z}_b^T, \mathbf{v}^T]^T$, where $B$ represents the number of frames of the magnitude spectrum $|Y|$. The pooling layer extracts a target-dependent utterance-level embedding from the target-dependent frame-level features. The utterance-level network is a regression network. It takes the utterance-level embedding as the input for predicting ${\rm{SNR}}^u$ of $y(t)$. It minimizes the mean-squared error:
\begin{equation}
J_1 = ||q - {\rm{SNR}}^u||^2_2
\end{equation}
where $q$ is an estimate of ${\rm{SNR}}^u$ (denoted as $\widehat{{\rm{SNR}}}^u$ in Figure \ref{fig:3}), which will be used as the channel weight for the channel-selection algorithm in the test stage.

The main network and auxiliary network are jointly trained by back-propagation. Both networks use mean pooling as the pooling layer, which averages the frame-level features along the time axis for the utterance-level embeddings.

\subsubsection{Channel selection algorithms}
 \begin{figure*}[t]
      \includegraphics[width=0.9\textwidth]{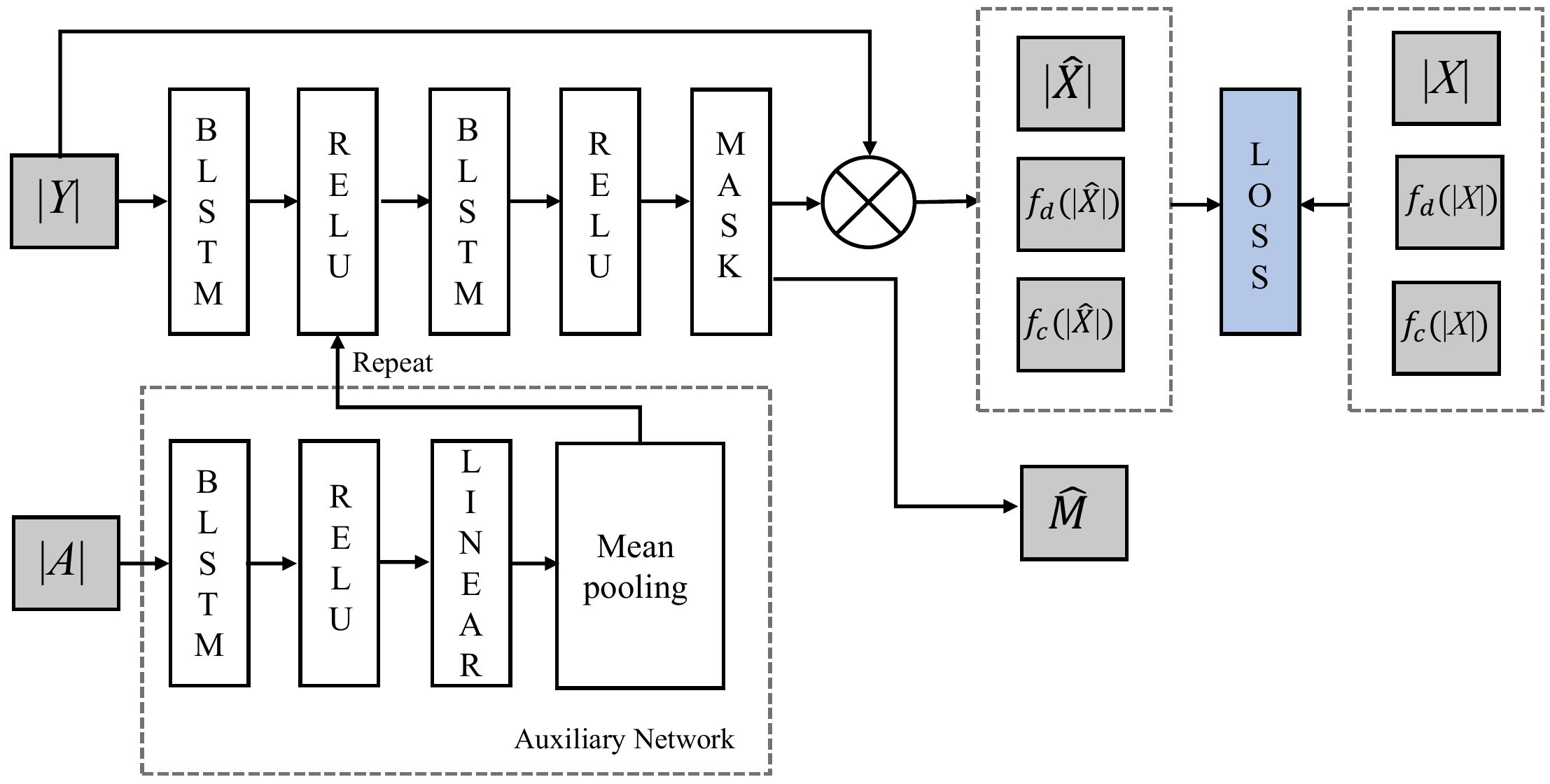}
  \caption{Diagram of the single-channel speaker extraction algorithm.}
    \label{fig:4}
\end{figure*}
The channel-selection algorithms in this section are used in the test stage only. Applying the channel-weight estimation network to each channel respectively gets a channel-weight vector $\mathbf{q}=[q_1,q_2,...,q_W]^T$. A channel-selection algorithm takes $\mathbf{q}$ as input, and outputs a channel-mask vector $\mathbf{p}=[p_1,p_2,...,p_W]^T$. Some channel-selection algorithms are described as follows.

\begin{itemize}
\item \textbf{Selecting one-best channel (1-best)}

   This algorithm selects the channel with the highest speech quality among the $W$ channels:
  \begin{equation}
   p_j=\left\{
   \begin{array}{ll}
    1, & {\mbox{if } q_j = \max_{1\leq n \leq w} q_n}\\
    0, & \mbox{otherwise}\\
   \end{array} \right.
   , \forall j=1,...W
  \end{equation}
  After the 1-best channel selection, a nonlinear single-channel speaker extraction algorithm will be applied.

\item \textbf{Selecting $N$-best channel with predefined number (fixed-N-best)}

  If the speakers are in a large room, and if the microphones are sufficiently dispersed, then selecting a number of microphones around the target speaker may yield better performance than using all microphones. This channel selection algorithm first sorts $\{q_1,q_2,...,q_W\}$ in descent order, denoted as $q_1'\geq q_2'\geq \ldots\geq q_W'$, and then picks the first $N$ channels with the highest $q$:
   \begin{equation}
   p_j=\left\{
   \begin{array}{ll}
    1, & \mbox{if } q_j \in \{q_1',q_2',..,q_N'\}\\
    0, & \mbox{otherwise}\\
   \end{array} \right.
   , \forall j=1,...W
  \end{equation}
  where $N\leq W$.

\item \textbf{Selecting $N$-best channel where number is predetermined on-the-fly (auto-N-best)}

  This algorithm provides a method to determine $N$ automatically. It first picks the 1-best channel by $q_*= max_{1\leq n \leq w}q_n$, and then calculates $p_j$ by:
  \begin{equation}
   p_j=\left\{
   \begin{array}{ll}
    1, & \mbox{if } \frac{q_j}{q_*}\frac{1-q_*}{1-q_j}>\gamma\\
    0, & \mbox{otherwise}\\
   \end{array} \right.
   , \forall j=1,...W
  \end{equation}
  where $\gamma \in [0,1]$ is a tunable hyperparameter.

\item \textbf{Selecting soft $N$-best channel (soft-N-best)}

  Different from the auto-N-best algorithm, this algorithm re-weights the selected channels according to the quality of the target speech:
  \begin{equation}
   p_j=\left\{
   \begin{array}{ll}
    q_j, & \mbox{if } \frac{q_j}{q_*}\frac{1-q_*}{1-q_j}>\gamma\\
    0, & \mbox{otherwise}\\
   \end{array} \right.
   , \forall j=1,...W
  \end{equation}

\end{itemize}

\subsection{Single-channel speaker extraction}\label{sec:3.2}

 The system diagram of the single-channel speaker extraction is shown in Figure \ref{fig:4}.
 Given some auxiliary information of the target speech, It estimates a ratio mask for the target speech. Then, the estimated magnitude spectrum of the target speaker is obtained by applying the ratio mask to the mixed speech as follows:
\begin{equation}
|\hat{X_{a}}(t,f)| = \hat{M_{a}}(t,f)|Y(t,f)|
\end{equation}
where $|\hat{X_{a}}(t,f)|$ is the estimated magnitude of the target speech, $ \hat{M_{a}}(t,f)$ is an estimated phase sensitive mask of the target speech.

  \begin{figure*}[t]
      \includegraphics[width=0.7\textwidth]{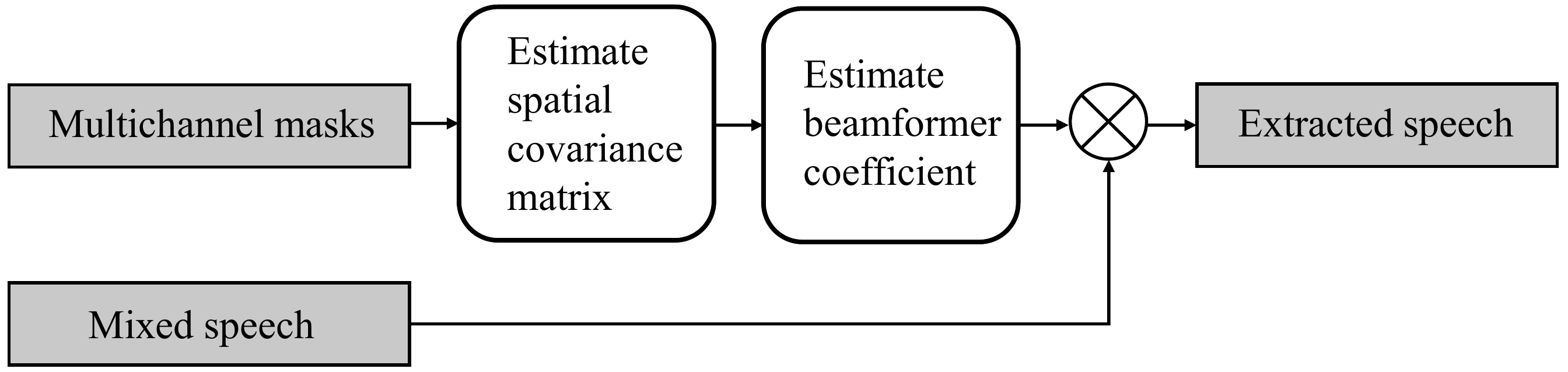}
  \caption{Diagram of beamforming.}
  \label{fig:5}
\end{figure*}

The single-channel speaker extraction encodes the auxiliary information of the target speaker into an embedding in the same way as that in the channel-weight estimation network. The network uses the magnitude and temporal spectrum approximation loss \cite{xu2019optimization}:
\begin{equation}
\begin{aligned}
J_2&=\frac{1}{T}\sum \Big(\big\||\hat{X_{a}}(t,f)|\\
&\qquad -|X(t,f)|\cos(\theta_{y}(t,f) -\theta_{x}(t,f))\big\|^2_F\\
   &+w_d\big\|f_d|\hat{X_{a}}(t,f)|\\
   &\qquad -f_d(|X(t,f)|\cos(\theta_{y}(t,f) -\theta_{x}(t,f)))\big\|^2_F\\
   &+w_c\big\|f_c|\hat{X_{a}}(t,f)|\\
   &\qquad -f_c(|X(t,f)|\cos(\theta_{y}(t,f) -\theta_{x}(t,f)))\big\|^2_F\Big)
\end{aligned}
\end{equation}
where $\theta_{y}(t,f)$ and $\theta_{x}(t,f)$ are the phases of the spectrum of the mixed speech and target direct speech, $w_d$ and $w_c$ are the weights, and $f_d(\cdot)$ and $f_c(\cdot)$ are two functions for calculating the increment and acceleration respectively \cite{1164788}. The loss function not only integrates the merit of the phase-sensitive mask \cite{erdogan2015phase} and signal approximation \cite{huang2014deep}, but also captures the dynamic information, i.e. the increment and acceleration, of the target speech.

\subsection{Beamforming algorithm}\label{sec:3.3}

As shown in Figure \ref{fig:5}, the deep learning based MVDR \cite{heymann2016neural} is used to integrate the selected channels for the target speech. It uses the single-channel speaker extraction in Section \ref{sec:3.2} to generate an estimation of the interference speaker in each channel, and then uses the estimations to learn linear filters $w_a$ from MVDR, and finally uses the linear filters to produce the target speech by:
\begin{equation}
\hat{X}_a(t,f) = w_a^H(f)Y(t,f)
\end{equation}
where the symbol $^H$ is the conjugate transpose operator, and $\hat{X_a}_(t,f)$ is the estimated target speech.

The filter is derived by:
\begin{equation}
w_a(f)=\frac{{\hat{\Phi}}^{-1}_{ii}(f)\hat{c}_a(f)}{\hat{c}_a^H(f){\hat{\Phi}}^{-1}_{ii}(f)\hat{c}_a(f)}
\end{equation}
where ${\hat{\Phi}}_{ii}(f)$ is an estimate of the spatial covariance matrix of the interference speaker, and $\hat{c}_a(f)$ is the first principal component of ${\hat{\Phi}}_{ii}(f)$:
\begin{equation}
{\hat{\Phi}}_{ii}(f) = \frac{1}{\sum_t \eta_{a} (t,f)}\sum_t \varepsilon_{a} (t,f)Y(t,f)Y(t,f)^H
\end{equation}
\begin{equation}
\hat{c}_u(f) = {\rm{principal}}({\hat{\Phi}}_{ii}(f))
\end{equation}
where $\varepsilon_a (t,f)$ is defined as the product of the estimated masks of the interference speaker obtained by the single-channel speaker extraction algorithm.
 \begin{equation}
\eta_{a} (t,f) = \prod_{i=1}^W \hat{M_{a}}(t,f)
\end{equation}

Finally, we can get the estimated target speech $\hat{x}_a(t)$ by inverse-STFT:
 \begin{equation}
\hat{x}_a(t) = \rm{iSTFT}(\hat{X}_a(t,f))
\end{equation}

\section{ Experiments}\label{sec:4}

In this section, we present the datasets, experimental settings, and results in Sections \ref{subsec:data}, \ref{subsec:settings}, and \ref{sec:5}, respectively.

\subsection{Datasets}\label{subsec:data}

We focused on the two speaker speech separation problem, in which one speaker was regarded as a target speaker.
For each mixture, we randomly generated 16 microphones in a randomly generated room that is 5 to 15 meters long, 5 to 25 meters wide, and 1 to 2.5 meters high. Microphones and speech sources were placed randomly in the room. Each speaker is at least 0.2 meter away from the walls, and 0.3 meter from the microphones. We created room impulse responses (RIRs) using the method in \cite{allen1979image}. We convolved the clean speech signals with the RIRs and added the reverberant signals from both sources to produce the mixes speech. The reverberant condition T60 was generated from a Gaussian distribution with a mean value of 0.25 second and a variance of 0.1 second. The largest T60 was limited to 0.8 second. To evaluate the effect of the number of microphones on performance, we repeated the above process except that the number of microphones was set to 8.

We generated a WSJ0-adhoc corpus from the WSJ0 corpus \cite{garofolo1993csr} at a sampling rate of 8 kHz in the aforementioned environment for 16 and 8 microphones respectively. The mixed speech in the WSJ0-adhoc corpus contains the same content as that in the WSJ0-2mix corpus \cite{xu2019optimization}. Specifically, the original corpus '$si\_tr\_s$', composed of 50 male and 51 female speakers, was randomly mixed to generate the training set (20000 utterances) and validation set (5000 utterances) of WSJ0-2mix at various SNR uniformly chosen between 0 dB and 5 dB. Similarly, the original data sets '$si\_dt\_05$' and '$si\_et\_05$', composed of 10 male and 8 female speakers, were randomly mixed to generate the test set (3000 utterances). Because the speakers of the test set were different from those of the training set and validation set, our experimental scenario was regarded as open condition evaluation.

For the mixed speech of two speakers, we set the first speaker as the target speaker, and the second speaker as the interference speaker. The utterance of the target speaker was regarded as the reference speech. At the same time, we randomly selected a different utterance of the same target speaker from WSJ0 corpus as the auxiliary information of the target speaker. To study the effect of different gender combinations on performance, we grouped the test set into 'Female $+$ Female' (F$+$F), 'Female $+$ Male' (F$+$M), 'Male $+$ Male' (M$+$M) for evaluation.

\subsection{Experimental settings}\label{subsec:settings}

\begin{table*}[t]
\caption{Comparison results of DABse with 16 microphones per ad-hoc microphone array and two baselines, where the average T60 of all environments is 0.25s.}
\label{tab:16ch}
  \begin{tabular}{cccccc}
    \hline
    Comparison method  & Beamforming method &Gender &SDR(dB)   & PESQ&STOI\\ \hline
      && M+M &2.84 & 1.91  &0.71\\
   Single-channel &-& F+F & 3.65 & 2.00&0.74\\
     & &F+M & 5.45  & 2.18&0.77\\ \hline
        && M+M &5.47 & 2.20 &0.79\\
    All-channels&MVDR& F+F & 4.17 & 1.92&0.78\\
     && F+M & 5.68  & 2.14&0.80\\ \hline
        && M+M & 4.84 & 1.98  &0.78\\
    DABse+1-best& -& F+F & 6.58 & 2.21&0.79\\
     && F+M & 7.72& 2.26&0.81\\ \hline
        && M+M & 3.51 & 1.86  &0.77\\

    DABse+fixed-N-best&MVDR & F+F & 5.32 & 2.08&0.79\\
     && F+M & 7.96  & 2.30&0.82\\ \hline
       & & M+M & 5.56 & 2.10  &0.79\\
    DABse+auto-N-best &MVDR& F+F & 6.66 & 2.23&0.81\\
     && F+M & 8.47 & 2.34&0.84\\ \hline
     & & M+M & 5.17 & 2.06  &0.76\\
    DABse+soft-N-best &MVDR& F+F &6.30 & 2.15&0.80\\
     && F+M & 8.11  & 2.30&0.83\\ \hline
  \end{tabular}
\end{table*}
\subsubsection{DABse}
We set the frame length to 32 milliseconds and the frame shift to 16 milliseconds. A 129-dimensional spectrum was extracted from each frame by STFT with a pre-emphasis of a normalized square root hamming window.

The parameters of the channel-weight estimation network were set as follows.
(i) For the auxiliary network, the BLSTM network contains 256 nodes in both the forward and backward directions. The feed-forward hidden layers contain a nonlinear layer of 256 rectified linear units (ReLU) and a linear layer of 30 nodes. Mean pooling is used as the pooling layer. Finally, we obtained a 30-dimensional embedding vector from the auxiliary utterance of the target speaker. (ii) For the main network, the magnitude feature of a mixture was used as the input of a BLSTM of 512 nodes in both directions. The activations of each frame were repeatedly concatenated with the target speaker embedding from the auxiliary network. The concatenated feature was uses as the input of another BLSTM of 512 nodes. A hidden layer with 512 ReLUs was built upon the BLSTM, followed by a mean pooling layer. Finally, a hidden layer with 256 ReLUs was built upon the mean pooling layer. The output layer was a 1-dimensional sigmoid function for predicting ${\rm{SNR}}^u$ of the target speaker.

 The learning rate was initialized by 0.0005 and scaled down by 0.7 when the training loss increased on the development set. The minibatch size was set to 32. The network was trained with at least 30 epochs and at most 60 epochs. Early stop was applied when the relative loss reduction between two successive epochs was lower than 0.01. The Adam algorithm was used to optimize the network.

 The parameters of the single-channel speaker extraction network were similar with those of the channel-weight estimation network, except that the output layer contains 129 units for predicting the ratio mask.

 We denote DABse with a specific channel-selection algorithm as 'DABse+channel-selection', which results in the following four methods.
 \begin{itemize}
\item \textbf{DABse+1-best.}
\item \textbf{DABse+fixed-N-best.}We set N at $\sqrt{M}$.
\item \textbf{DABse+auto-N-best.}We set $\gamma$ at 0.5.
\item \textbf{DABse+soft-N-best.}We set $\gamma$ at 0.5.
\end{itemize}
 Note that 'DABse+1-best' is a nonlinear speech separation method while the others are linear methods.

\subsubsection{Baselines}
We compared DABse with two extreme DABse variants:
\begin{itemize}
\item \textbf{Selecting one-random channel (single-channel)}
  We randomly select a channel from the $W$ channels, and then conduct single-channel speaker extraction. This extreme case does not refer to channel selection, and is therefore irrelevant to the number of channels.

  \item \textbf{Selecting all channels (all-channels)}
  We use all channels for the multi-channel speaker extraction, i.e.:
  \begin{equation}
 p_j=1,\forall j=1,...W
    \end{equation}
\end{itemize}
 Note that 'single-channel' is a nonlinear speech separation method while 'all-channels' is a linear method.

\subsubsection{Evaluation metrics }
The performance evaluation metrics include signal to distortion ratio (SDR) \cite{vincent2006performance}, perceptual evaluation of speech quality (PESQ) \cite{rix2001perceptual}, and short-time objective intelligibility (STOI) \cite{taal2011algorithm}. SDR is a metric similar to SNR for evaluating the quality of enhancement. PESQ is a test methodology for automated assessment of the speech quality as experienced by a listener of a telephony system. STOI evaluates the objective speech intelligibility of time-domain signals. The higher the value of an evaluation metric is, the better the performance is.

\subsection{Results}\label{sec:5}

Table \ref{tab:16ch} lists the comparison results of the DABse variants with the two baselines. From the table, we see that the last three channel selection algorithms outperform the other comparison methods in most cases. Among the three algorithms, 'DABse+auto-N-best' outperforms all comparison methods, followed by 'DABse+soft-N-best'. Although 'DABse+fixed-N-best' produces a similar PESQ result with 'DABse+soft-N-best', its SDR and STOI scores are poorer than the latter.

From the table, we also find that even the simplest channel selection algorithm 'DABse+1-best' can produce better experimental results than both 'single-channel' and 'all-channels', which proves the necessity and effectiveness of the channel selection for DABse. For example, for the combination of F$+$M, 'DABse+1-best' achieves 2.27dB higher than 'single-channel' in terms of SDR.

\subsection{Effect of the number of microphones in the ad-hoc microphone array}

In order to explore the influence of different number of microphones of ad-hoc microphone arrays on DABse, we conducted experiments with ad-hoc microphone arrays of 8 microphones on the F+M combination when $T60_{mean}$ was set at 0.25 seconds. And it is necessary to be mentioned that we set N at 3 of 'DABse+fixed-N-best'.

The results are listed in the Table \ref{tab:8ch}. From the table, we can find that it is the 'DABse+soft-N-best' rather than 'DABse+auto-N-best' that performs the best among all comparison methods. Besides, it is obvious that the ad-hoc microphone array with 16 microphones outperforms that with 8 microphones.
\begin{table}[t]
\caption{Comparison results of DABse with 8 microphones per ad-hoc microphone array and two baselines on the gender pair of F$+$M.}
\label{tab:8ch}
  \begin{tabular}{ccccc}
    \hline
   Comparison & Beamforming  & SDR  & PESQ   & STOI\\
   method&         method   &     (dB)      &         &     \\\hline
    Single-channel &-& 5.45  & 2.18&0.77\\\hline
    All-channels &MVDR&3.52  & 1.94   & 0.74\\\hline
    DABse+1-best &-& 5.69 & 2.05 & 0.76\\\hline
    DABse+fixed-N-best &MVDR&  5.79 &2.28   &0.77 \\\hline
    DABse+auto-N-best &MVDR& 5.84 & 2.17   &0.79 \\\hline
    DABse+soft-N-best &MVDR& 6.49 & 2.22   &0.80 \\ \hline
  \end{tabular}
\end{table}

\subsection{Effect of DABse on different gender combinations. }
 Table \ref{tab:16ch} lists the effect of DABse on different gender combinations. From the table, we see that DABse and single-channel speaker extraction always achieve better performance on the gender pair of F$+$M. For the same gender combinations, it seems that they perform better on F$+$F than on M$+$M in most cases. The phenomena indicate that the effectiveness of DABse is strongly affected by the single-channel speaker extraction algorithm.

\subsection{Effect of hyperparameters of DABse on performance. }

\subsubsection{DABse+fixed-N-best}
To study how the selected number of channels $N$ affects the performance of 'DABse+fixed-N-best', we conducted an experiment on the F$+$M gender pair with $N$ selected from $\{2,4,6,8,10,12,14, 16\}$ respectively. From the experimental results in Table \ref{tab:N}, we see that
 the performance first gets improved and then decreased along with the increase of $N$, with the top SDR performance appearing at $N=4$ and top STOI performance appearing at $N=6$, which demonstrates the correctness of our experimental setting.
\begin{table}[t]
\caption{Effect of hyperparameter $N$ of 'DABse+fixed-N-best' on the gender pair F$+$M.}
\label{tab:N}
  \begin{tabular}{cccc}
    \hline
   Number of  & SDR  & PESQ   & STOI\\
   selected microphones(N)&(dB) & & \\\hline
   N = 2 & 4.42 & 2.11 & 0.78\\
   N = 4 & 7.96& 2.30  & 0.83\\
   N = 6 &7.87  & 2.30   & 0.85\\
   N = 8 & 7.53 & 2.27   & 0.83\\
   N = 10 & 6.97  & 2.24   & 0.81\\
   N = 12 & 6.23&  2.22  & 0.81\\
   N = 14 & 6.76 & 2.26  & 0.82\\
   N = 16 & 5.86 & 2.14  & 0.80\\ \hline
  \end{tabular}
\end{table}

\subsubsection{DABse+auto-N-best and DABse+soft-N-best. }

\begin{figure}[h]
      \includegraphics[width=0.47\textwidth]{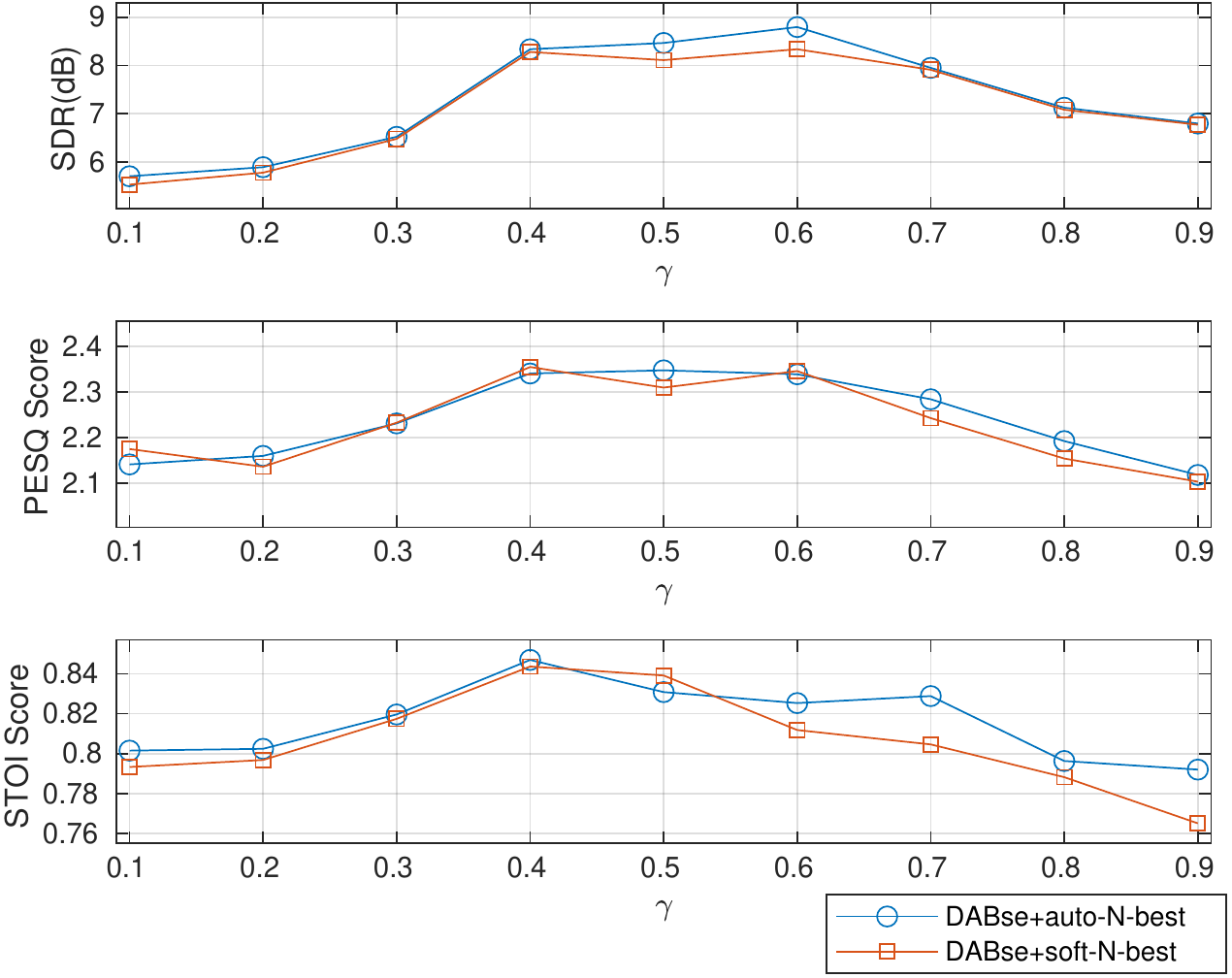}
  \caption{Effect of hyperparameter $\gamma$ of 'DABse+auto-N-best' and 'DABse+soft-N-best' on the gender pair F$+$M.}
  \label{fig:6}
\end{figure}

To explore how the hyperparameter $\gamma$ affects the performance of 'DABse+auto-N-best' and 'DABse+soft-N-best', we conducted an experiment on the F$+$M gender pair with $\gamma$ selected from 0.1 to 0.9. The experimental results are shown in Figure \ref{fig:6}. From the figure, we see that, when $\gamma$ was set at 0.4 to 0.6, both 'DABse+auto-N-best' and 'DABse+soft-N-best' achieve top performance. Therefore, setting the default value of $\gamma$ to 0.5 is reasonable.

\section{Conclusions}\label{sec:6}

In this paper, we have proposed deep ad-hoc beamforming based on speaker extraction, which is the first work of the target-dependent speech separation based on ad-hoc microphone arrays and deep learning. DABse uses the channel-weight estimation network based on speaker extraction to estimate the SNR of the target speaker, and then takes the SNR as the channel weight for the selection of high-quality channels, and finally takes the selected channels for the deep learning based MVDR. The deep learning based MVDR first takes the single-channel target-dependent speaker extraction network to estimate the clean spectrum of the target speech at each selected channel, and then uses the estimated spectrum to derive an MVDR filter for the final speech separation. Because the two deep models in DABse are trained in a single-channel fashion, it is able to handle any number of microphones in the test stage. Because MVDR is a linear filter, DABse does not suffer from nonlinear distortions.
We have conducted extensive experiments in the scenarios where the speech sources are located randomly in large rooms. We compared DABse with the baselines of 'single-channel' and 'all-channels'. Experimental results demonstrate that the proposed DABse outperforms the baselines significantly, which illustrates the effectiveness of DABse in the adverse environments.

\begin{backmatter}

\section*{Acknowledgements}
The authors would like to thank Zhongxin Bai and Junqi Chen for helpful discussions.

%
%
%

\section*{Competing interests}
The authors declare that they have no competing interests.

\section*{Consent for publication}
All authors agree to the publication in this journal.

\section*{Authors' contributions}
Model development: ZY, SG ,and XZ.
Experimental testing: ZY and SG .
Writing paper:ZY, XZ ,and SG .
The authors read and approved the final manuscript.

\section*{Authors' information}
Text for this section\ldots


\bibliographystyle{bmc-mathphys} 
\bibliography{referrence}      









\end{backmatter}

\end{document}